\documentclass[conference]{IEEEtran}
\usepackage{blindtext, graphicx}
\usepackage{cite}

\ifCLASSINFOpdf
\else
\fi
\usepackage{url}

\usepackage{soul}
\usepackage{color}

\begin{document}
%
\title{Towards a Foundation for Holistic Power System Validation and Testing}

\author{\IEEEauthorblockN{M. Blank, S. Lehnhoff}%
\IEEEauthorblockA{R\&D Division Energy\\%
OFFIS -- Institute for Information Technology\\%
Oldenburg, Germany\\%
\footnotesize{\{marita.blank, sebastian.lehnhoff\}@offis.de}}%
\and
\IEEEauthorblockN{K. Heussen, D. E. Morales Bondy}%
\IEEEauthorblockA{Department of Electrical Engineering\\%
Technical University of Denmark\\%
Kgs. Lyngby, Denmark\\%
\footnotesize{\{kh, bondy\}@elektro.dtu.dk}}%
\and
\IEEEauthorblockN{C. Moyo, T. Strasser}%
\IEEEauthorblockA{Energy Department\\%
AIT Austrian Institute of Technology\\%
Vienna, Austria\\%
\footnotesize{\{cyndi.moyo, thomas.strasser\}@ait.ac.at}}}

\IEEEoverridecommandlockouts 


%


\maketitle

\begin{abstract}
Renewable energy sources and further electrification of energy consumption are key enablers for decreasing greenhouse gas emissions, but also introduce increased complexity within the electric power system.
The increased availability of automation, information and communication technology, and intelligent solutions for system operation have transformed the power system into a smart grid. 
In order to support the development process of smart grid solutions on the system level, testing has to be done in a holistic manner, covering the multi-domain aspect of such complex systems.
This paper introduces the concept of holistic power system testing and discuss first steps towards a corresponding methodology that is being developed in the European ERIGrid research infrastructure project.
\end{abstract}



%
\IEEEpeerreviewmaketitle

\section{Introduction}\label{sec:introduction}
Efforts to reduce green-house gas emissions have already had a strong effect on the power system. Integration of renewable energy sources (RES), flexible loads and storage systems into the power system has increased steadily over the past years. 
This has introduced challenges to power system operators due to the fluctuating nature of RES, increased complexity in the system, and heterogeneous components \cite{SET}.
The increased implementation of advanced automation and information and communication technologies (ICT) 
are transforming the power system into a cyber-physical system which integrates infrastructures of different domains -- a smart grid \cite{SmartGridsPath, SmartGridsRoadmap}.
As such, it is necessary to implement integrated solutions for operating the system that fulfil high-reliability, real-time or regulatory requirements, just to name a few.
Before deployment such solutions have to be validated and tested.
Until now, only certain aspects with a main focus on components are tested \cite{bruendlinger2015}.
However, in order to support the different stages of the overall development process for smart grid solutions, tests in a holistic manner are needed, i.e. integrating different domains on system level \cite{CIGRE}.

The project ERIGrid\footnote{https://www.erigrid.eu/}
addresses the challenging aims of a holistic system testing approach for smart grids 
by creating a platform and methodology for integrating 18 European research centres and institutions.
A holistic testing methodology allows tests and experiments representative of integrated smart grids to be conducted through testing and experimentation across distributed research infrastructures (RI) which may not necessarily be functionally interconnected.

The paper is outlined as follows.
In Section~\ref{sec:concept}, the concept of holistic testing and its advantages are briefly summarised. 
In Section~\ref{sec:approach} the general approach of the ERIGrid holistic testing concept is presented. 
In Section~\ref{sec:results} the current status and preliminary results on state of the art and first concepts for test case description are presented, which are explained on an example in Section~\ref{sec:example}. 
The paper finishes with the conclusions in Section~\ref{sec:conclusion}.

\section{Concept of Holistic Testing}\label{sec:concept}
A holistic smart grid research and development approach not only addresses the whole development cycle (design, analysis, simulation, experimentation, testing and deployment), but must also take into account all relevant components, facets, influences that future power systems will comprise of, all of which may affect the controller, algorithm(s), or use case in question.
Testing highly integrated systems without taking into account possible disturbances by users, markets, ICT availability, etc, is invalid. Formal analysis of these vastly complex, integrated systems is not (yet -- if at all) possible. Hence, rigorous testing strategies are required that allow for the validation of integrated systems of different domains represented at different RI. Due to the importance of the system at hand and the immaturity of controllers, applications, and hardware, real-world embedded field tests are, in many cases, out of the question.
Although a functional integration of the aforementioned RI running in parallel and yielding integrated holistic energy systems is theoretically possible it remains practically infeasible.
In order to be capable of conducting tests and experiments representative of integrated smart grid systems, testing and experimentation must be possible across distributed and not necessarily functionally interconnected RI.

The outcomes of experiments at different RI are dependent on each other and must be analysed in an integrated way. 
ERIGrid proposes an approach for realizing a 
holistic procedure for smart grid system validation to support comparability between experiments of different setup and design, 
thus facilitating subsequent re-utilization of experimental results from different stakeholders through consecutive, 
sequential, and parallel experiments. 

In the following section a cyber-physical energy system based procedure for holistic testing is proposed.

\section{Towards a Holistic Testing Procedure}\label{sec:approach}
\begin{figure}[!t]
\centering
\includegraphics[width=0.95\columnwidth]{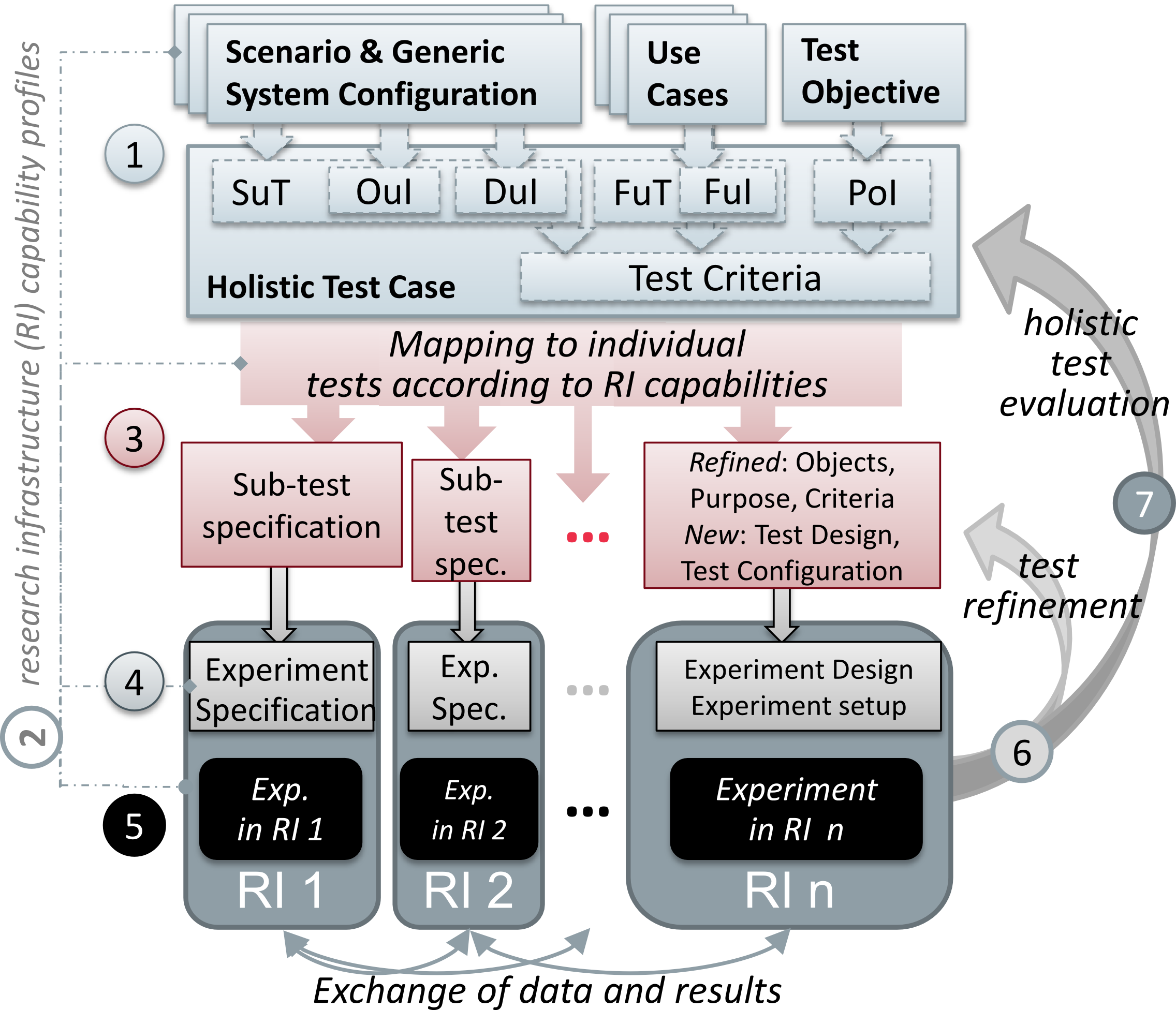}
\caption{Descriptive elements in a holistic test specification.
Abbreviations in Step~1 are explained in Section~\ref{sec:holistic_test_definition}.}
\label{fig:holistic_testing_revised}
\end{figure}

This section details the approach within ERIGrid to realise holistic testing. 
Fig.~\ref{fig:holistic_testing_revised} illustrates the main steps.

\subsection{Holistic Testing Steps}
The starting point of the envisioned procedure is at the specification of a \emph{holistic test case} (i.e., Step~1). This is derived from a scenario, a corresponding system configuration and the use cases within this setup. Consequently,
the test case is positione to identify specific test criteria, relating to the test system configuration, relevant use cases and a specific test objective. 
In an independent step, the RI involved are profiled with regards to their testing capabilities (i.e., Step~2).
As mentioned above, the procedure assumes that for such a holistic test it is not feasible to define and conduct a combined large-scale test incorporating all relevant domains and systems in one single setup. Therefore, the holistic test must be divided into sub-tests.
The sub-tests concentrate on certain components or sub-systems in total reflecting the structure of the holistic test in such a way that the sub-test results may be assembled to offer quantitative feedback on the holistic test criteria. This decomposition is performed in the first part of the mapping step (i.e., Step~3), where the interfaces and dependencies between the sub-test cases as well as the resulting requirements must be specified as well. 

In the second part of the mapping step, the descriptions of the sub-test cases, considering the RI profiles from Step~2, are used to identify the appropriate RIs capable of conducting the test for each sub-test case. 
Once the RI and tests are known the experiments can be specified, e.g., the concrete setup and design (i.e., Step~4).
Within the context of carrying out the sub-tests (i.e., Step~5) it is necessary to analyse and to exchange data and results (i.e., Step~6) between the sub-tests, based on which cross-dependencies have been identified in Step~3.
The results of all tests are analysed and combined to obtain the criteria with which the holistic test is evaluated (i.e., Step~7). Possible methods for combining results might be up-scaling or aggregating results.
Thus, the mapping between the test has two purposes: \emph{(i)} the re-use of results as an input to generate successive results, and \emph{(ii)} the combination of results from different sub-tests to obtain results of the holistic test.
To this end, dependencies between tests have to be considered beforehand.
The mapping step as well as the step of combining results of the sub-test might be an iterative approach. Before setting up and conducting the experiments the process from holistic test to RI and back should be specified as precisely as possible to minimise the effort and costs.

\subsection{Holistic Test Case Specification}
\label{sec:holistic_test_definition}
A test specification aims to clarify the object under investigation, test objective, and by what means a test is to be carried out (i.e., test setup and test design): \emph{(i)} what needs to be tested, \emph{(ii)} why, and \emph{(iii)} how.  
As outlined above, the holistic testing procedure envisions a separation of the first two pillars of a test specification (i.e., test object and test objective) from the third (the means of testing). We refer to a holistic \emph{test case} as specification of the what and why of a test, without including specific limitations on test setup and test design\footnote{This kind of definition corresponds to the ICT test case definition \cite{baker2007}}.
In contrast to conventional power systems testing, this requires a more formal approach, as the intention of a test case must be unambiguously identifiable, enabling specification of a test design, and test setup in a separate step (see Section~\ref{sec:common_meta-description}).  

Another aspect of the holistic testing approach is the merger of different cultures of testing, which can be portrayed as a device-oriented culture of physical testing and a culture of testing ICT objects such as implementations of protocols and algorithms.
Rigorous formal specification of test cases as well as automated execution of tests are common in the ICT domain \cite{gnesi2012}. 
In the testing of physical components, the test object is delimited by its physical boundaries, requiring little further formalization of the test object. However, a good test specification requires insight on physical and engineering principles. Test specifications therefore tend to be domain specific and less formal. Further, much of the test design is decided by the available test setup. 
A challenge is therefore to formalize the complete cyber-physical system context and test criteria, to formulate a test case combining several ICT and physical components and sub-systems as well as test criteria spanning different domains.  

As illustrated in Fig.~\ref{fig:holistic_testing_revised}, we envision the specification of a \emph{holistic test case} as composed of the following description items: 
Given a smart grid scenario composed of a \emph{systems configuration} (SC) and related \emph{use cases}, as well as the intention of a \emph{test objective}, the test case intention is summarized in a  \textbf{Narrative}.
With reference to the SC, the \emph{System under Test} (\textbf{SuT}) identifies the system boundaries of an abstract test setup entailing all relevant interactions requiring investigation, and the \emph{Object under Investigation} (\textbf{OuI}) identifies to the system, subsystem or component
 with respect to which the \emph{test criteria} will be formalized. 
The \emph{Domain(s) under Investigation} (\textbf{DoI}) identify the relevant physical or cyber-domains of test parameters and connectivity. 
With reference to use cases, the full set of \emph{Function(s) under Test} (\textbf{FuT}), and the specific \emph{Function under Investigation}  (\textbf{FuI}) are identified. 
The \emph{Purpose of Investigation} (\textbf{PoI}) formulates the test objective, also stating whether it relates to \emph{characterization, validation or verification} objectives. 
Together the above items inform the 
\textbf{Test Criteria}, which formalize the test metrics into \emph{target criteria}, \emph{variability attributes}, and \emph{quality attributes} (thresholds). 
An example of such a test case specification is provided in Section~\ref{sec:example}.

\subsection{Towards a Common Meta-Description for Sub-Tests}\label{sec:common_meta-description}

In order to support the mapping process, a common meta-description of tests is needed. This facilitates the three parts of the mapping introduced above: 
\emph{(i)} the mapping of the holistic test to sub-tests,
\emph{(ii)} the mapping of sub-tests to RI, and
\emph{(iii)} the mapping between tests reflecting the interdependencies between them. The latter refers to (re-)using results from one test in another test.
To this end, a methodology is proposed to classify tests with regard to different categories. 
For each category a classification is identified from existing testing procedures that summarizes or aggregates tests with common properties.
These categories cover different information needed for specifying sub-tests including information given in the holistic description (e.g., PoI, OuI) but also detailed information such as test setup and test design.

There are relations between the categories defining a hierarchy or order between them.
For instance, the purpose of investigation of a test determines the test criteria to be investigated. Once relations between categories have been identified dependencies between particular classes of different categories can be analysed, 
i.e., which class from one category can be combined with which class from another category.
Firstly, this information can be used for specifying requirements within a test, i.e., information needed for the experiment specification of e.g., the design, objects, and (sub-)system to be considered within a test.
Secondly, the information can be used between tests for realising a holistic test as a combination of sub-tests. From the holistic test, requirements on sub-tests have to be specified in terms of the given categories as well as requirements on information exchange between sub-tests.
Given this categorisation and requirements the sub-tests can be derived and specified.
Thirdly, if the RI profiling is mapped to the same categories this facilitates the choice for RI capable for certain tests and thus the mapping to RI.

Perspectively, a common, formal meta-description and derived rules about relations, e.g. in form of an ontology, can support the mapping steps mentioned above. With that, conclusions can be drawn on a given test specification, e.g., if information is missing within the specification.
Furthermore, an automated decomposition of a holistic test into individual test and RI might be enabled.

\section{Preliminary Results}\label{sec:results}
In this section, first results of the work in progress are presented that aim towards the realisation of the introduced validation and testing approach.

\subsection{Assessment of Current Practice}
Information about tests that are currently conducted at RI of all ERIGrid partners have been collected with the help of a questionnaire in order to derive a first classification of tests as proposed in Section~\ref{sec:common_meta-description}.
The information have been clustered with regard to the following different categories:
\emph{(i)} purpose of investigation, \emph{(ii)} test setup, \emph{(iii)} test criteria, \emph{(iv)} test design, and \emph{(v)} object of investigation.
The work has been separated into five working groups each responsible for one of the categories. The aim has been to identify different types of tests from the questionnaires as clusters and give a description of the common properties.
After this initial activity, another working group has been initialised for aligning the results of the five working groups regarding definitions and relations between clusters. 
The result has been a first consolidated description for classes of the categories.

It has been identified that additional information and categorisation is necessary on interfaces or connections of one test to other components or (sub-)systems. This will facilitate the mapping between different sub-tests.
In a next step, the tests covered in the questionnaires will be sorted to the classes in order to obtain a first impression about relations between classes of different categories.
Giving a holistic test specification, the classification can be adapted and refined and relations between classes and categories be detailed for classes that are relevant to this specific holistic test.

\subsection{Description of a Cyber-Physical System Configuration}\label{subsec:testcasedescription}
A cyber-physical system configuration in a smart grid context comprises physical components and devices, as well as various forms of ICT objects and relevant abstract components (e.g., markets, services). A continuum between concrete and abstract objects and their interconnections needs to be formally represented to specify a holistic (multi-domain) system under test. We define systems, domains, components, etc. as illustrated in Fig.~\ref{fig:examplediagram}; definitions wrt. standards / lexical terms.

\subsection{Related Work on System Testing}
The testing objectives in a holistic testing procedure should be viewed in context of a systems design procedure, as the goals and conditions of a test vary at different stages of development. For example during an early stage of systems development, a test may aim at the characterization and algorithm's performance, whereas at a later, more mature, stage also conformance with specific standards may be required. The view of system maturity and corresponding testing needs are outlined in different development models (integrated specification development and testing), such as test-driven, or agile development in ICT domain, or the well-known V-model or the W-model \cite{beizer1990}. 

\section{Example of Test Case Specification}\label{sec:example}
\begin{figure}[!t]
\centering
\includegraphics[width=0.75\columnwidth]{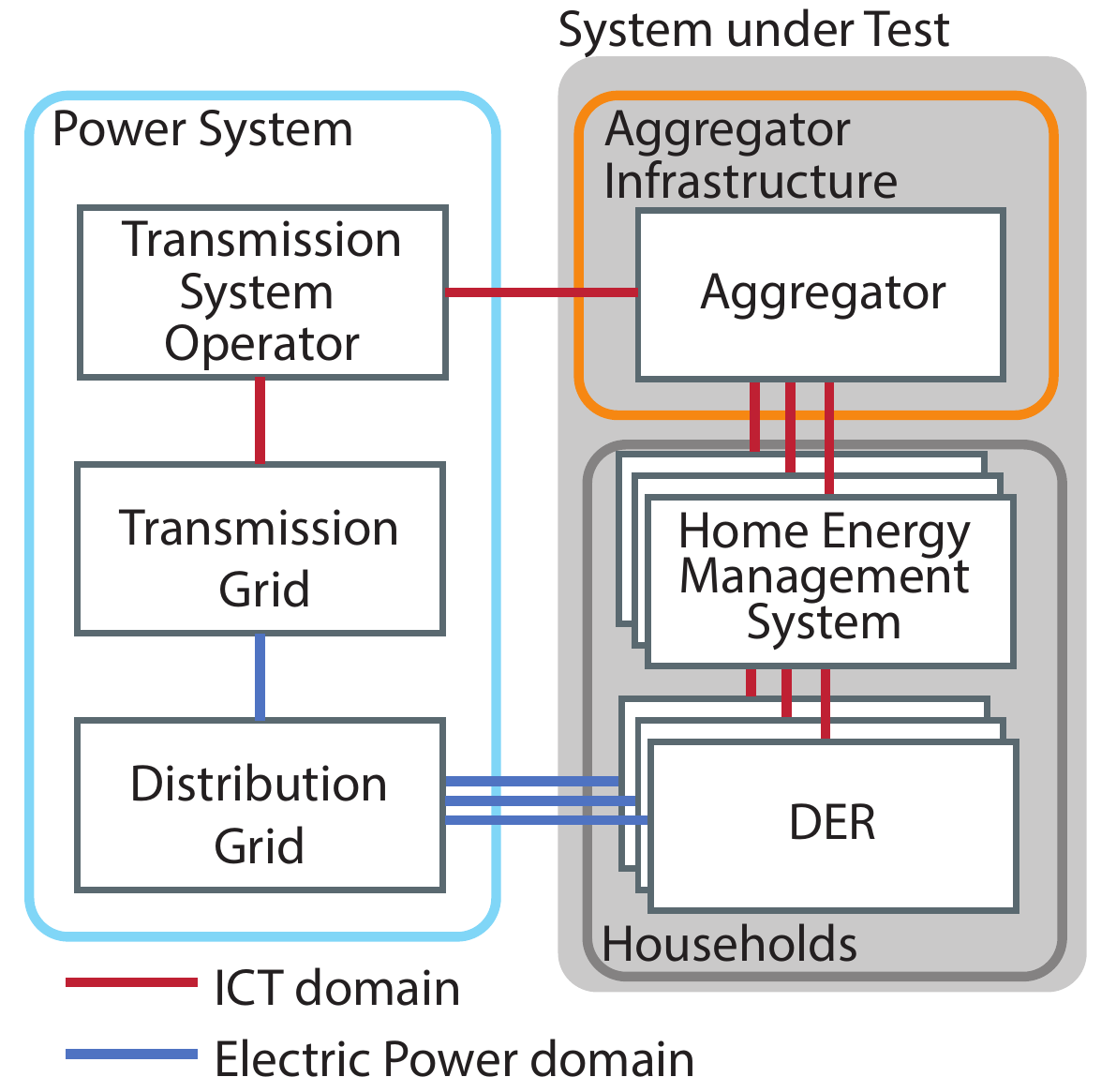}
\caption{The test setup described by the component centric approach. Note that the aggregator infrastructure and households form the \textbf{SuT}.}%
\label{fig:examplediagram}
\end{figure}

To exemplify a \emph{holistic test case} specification (see Section~\ref{subsec:testcasedescription}), an example is mapped to the holistic description:

\textbf{Narrative:} 
An aggregator, controlling DERs in 500 households, wishes to participate in the ancillary service markets by providing secondary frequency control to the transmission system operator (TSO). The presented example is a part of pre-qualification tests an aggregator must pass \cite{bondy2016procedure} in order to participate in aforementioned markets. Fig.~\ref{fig:examplediagram} presents the general system configuration. In this test case we analyse how the aggregator control system tracks the Automatic Generation Control (AGC) signal supplied by the TSO when subjected to disturbances in its ICT infrastructure. Metering issues and impact on the distribution grid are out of scope of this specific test. The holistic test case is described by:
\begin{itemize}
    \item \textbf{PoI:} to characterize the sensitivity towards ICT disturbances of the ancillary service quality of an aggregator.
    \item \textbf{SuT:} the system under test is composed of the aggregator infrastructure and 500 households. The input to the system is the AGC signal sent by the TSO and the output is the power consumption/production of the households.
    \begin{itemize}
    \item \textbf{OuI:} aggregator control system (part of the aggregator infrastructure).
    \item \textbf{DuI:} electric power and ICT infrastructure.
    \end{itemize}
    \item \textbf{FuT:} Aggregator central control, local DER control, communication functionality between aggregator and home energy management system (HEMS).
    \begin{itemize}
        \item \textbf{FuI:} Aggregator central control.
    \end{itemize}
    \item \textbf{Test Criteria:} 
    \begin{itemize}
        \item \emph{Target criteria:} Service quality, measured as the difference between the reference signal from the TSO and the aggregated power consumption/production of the DERs as measured by the individual DER measurement systems.
        \item \emph{Variability Attributes/Test Factors:} The ICT connection between aggregator and HEMS.
        \item \emph{Quality Attributes/thresholds:} The ICT parameters are to be varied until the aggregator is unable to track the AGC according to the contract.
    \end{itemize}
\end{itemize}

A sub-test could be a set of physical tests for an aggregator--single household setup, which has as outcome an availability/disturbance model to be used in another (simulation) test setup including a 500 household controller hardware-in-the-loop simulation.

\section{Conclusions}\label{sec:conclusion}
This paper has presented the concept of holistic power system testing to allow for an integrated test of smart grid solutions conducted in different RI.
An approach towards a holistic testing procedure has been proposed.
First results of the approach have been outlined and an exemplification of the outcome given.

In future work, the concepts and definitions will be adapted and refined. The methodological work is strongly interlinked with practical realisation of a holistic testing system and results and experiences will be taken into account in specifying the holistic testing procedure.
A core element is the development of the mapping process from a holistic test to sub-tests and RI that will build on the work presented here.
Concepts from design of experiments will be investigated for coupling results of tests from different RI.

\section*{Acknowledgment}

This work is supported by the European Community's Horizon 2020 Program (H2020/2014-2020) under project ``ERIGrid'' (Grant Agreement No. 654113). Further information is available at the corresponding website www.erigrid.eu. 

The authors would also like to thank all ERIGrid team members contributing to the development so far.

\ifCLASSOPTIONcaptionsoff
  \newpage
\fi

\end{document}